%% file: CSCW/main.tex
\documentclass[preprint]{acmart}
\AtBeginDocument{%
  }

\setcopyright{acmlicensed}
\copyrightyear{2026}
\acmYear{2026}
\acmDOI{XXXXXXX.XXXXXXX}
\acmConference[CSCW '26]{ACM SIGCHI Conference on Computer-Supported Cooperative Work \& Social Computing}{October 2026}{USA}
\acmISBN{978-1-4503-XXXX-X/2026}

\usepackage{enumitem}
\usepackage{hyperref}
\usepackage{multirow}
\usepackage{makecell}
\usepackage{longtable}
\usepackage{array}

\usepackage{tcolorbox}
\usepackage{microtype}

\begin{document}


\title{Federating Governance: How Community Rules Scale with Mastodon Instances}


\author{Rasika Muralidharan}
\email{rasimura@iu.edu}
\affiliation{%
  \institution{Luddy School of Informatics, Computing, and Engineering, Indiana University}
  \city{Bloomington}
  \state{Indiana}
  \country{USA}
}

\author{Yong-Yeol Ahn}
\email{yyahn@virginia.edu}
\affiliation{%
  \institution{University of Virginia}
  \city{Charlottesville}
  \state{Virginia}
  \country{USA}
}

\author{Bao Tran Truong}
\email{baotruon@iu.edu}
\affiliation{%
  \institution{Center Synergy of Systems, TUD Dresden University of Technology}
  \city{Dresden}
  \country{Germany}
}

\titlenote{Our repository, including code and data, is available at
\url{https://github.com/Rasikamurali/MastodonModeration}.}


\renewcommand{\shortauthors}{Muralidharan et al.}

\begin{abstract}

The rise of decentralized social media platforms like Mastodon and Bluesky highlights the challenge of scaling self-governance and moderation. As communities grow, they face new issues that demand increasingly complex governance structures. However, as moderation is mainly volunteer-driven, there is limited formal guidance on how community rules and moderation practices should evolve with growth. This study investigates how moderation scale with Mastodon instances by analyzing community rules across servers of varying sizes. We categorize these rules to identify key governance priorities and find that these priorities are remarkably consistent across instance sizes: rules addressing problematic content, such as harassment, hate speech, and illegal content, dominate regardless of scale. While smaller communities focus on narrower sets of topics, larger servers maintain a more balanced coverage of a broad range of topics. Our analysis of rule formalization reveals that community size strongly predicts rule development. As instances grow, their rules become more extensive and topically diverse, but also exhibit lower readability and linguistic diversity. In contrast, external federation interactions have a limited role, mainly associated with a broader scope of rules without substantially affecting their diversity or form. These findings highlight the relative influence of internal versus external factors, suggesting that local scaling pressures outweigh network-level dynamics in decentralized social media governance. The scaling pattern observed on Mastodon resemble those previously identified on centralized platforms such as Reddit, suggesting that community size imposes fundamental constraints on self-governance that transcend platform architectures.

\end{abstract}




\maketitle
\section{Introduction} \label{Introduction}

Balancing platform safety with constructive participation is a longstanding and unresolved problem in social media governance \cite{gillespie_custodians_2018, seering_moderator_2019, seering_reconsidering_2020}.
Major centralized platforms such as X \cite{twitter2025transparency}, YouTube \cite{youtube2024}, and Meta \cite{meta2024transparency} increasingly depend on automated moderation systems built around globally defined policies that are enforced uniformly across diverse user populations.  
These systems tend to focus on ostensibly universal content categories---such as unlawful content, incivility, hate speech, and harassment---that are assumed to be universal across all users and communities.
While this approach enables moderation at scale, it privileges mainstream norms and may overlook diverse cultural norms, values, and forms of expression used by minority and marginalized groups \cite{lingel2015face, jiang2020characterizing, griffin2024heteronormative, celeste2023platform}.

In response to these limitations, decentralized social media platforms such as Mastodon and Bluesky have gained prominence, renewing interest in community self-governance. These platforms distribute authority across independently operated servers (i.e., ``instances''), each self-governed by locally defined rules. Instances coexist within a network, forming a polycentric governance model that allows both local autonomy and cross-community interactions. Instances may choose to interoperate (i.e., ``federating''), restrict, or sever connections with one another (i.e., ``defederation''). As a condition of federation, communities agree to a minimal set of basic norms, such as prohibitions of hate speech, a model characterized by prior work as ``covenantal federalism'' \cite{jhaver_decentralize_platpower}. 
While decentralized governance addresses some of the shortcomings of centralized platforms---particularly their failure to account for contextual and diverse norms~\cite{DiazHechtFelella2021}---it does not resolve scalability, which remains a central concern for moderation~\cite{gillespie_content_2020}. 

As a community grows, moderators are faced with not only increased volumes of content, but also more diverse sets of users and problematic behaviors \cite{huang_decentralized_2024, lin_better_2017}. At the same time, they need to coordinate within larger moderator teams~\cite{bhattacharya2024unveiling}.
Even on comparatively well-resourced centralized platforms, moderators report burnout, inconsistency in enforcement, and the need for additional automated support. Studies of moderators on Facebook Groups~\cite{kuo2023unsung}, and Reddit moderators ~\cite{chandrasekharan2019crossmod, dosono_moderation_2019, hill_how_2019, lloyd_there_2025} document these challenges. On Mastodon, these challenges are amplified as moderation is almost entirely volunteer-driven. Moderation workloads thus grow without proportional increases in monetary or organizational resources such as tools or standardized training. This creates a natural setting to test whether governance scaling patterns observed in centralized platforms---which can invest in moderation infrastructure as they grow---also emerge in systems where communities must adapt independently.

Federation further complicates moderation at scale. Beyond enforcing local norms, effective moderation in federated networks requires balancing these norms with those introduced through federation~\cite{defederation2025}. Each instance must continually articulate and renegotiate existing and emerging norms, as users routinely participate across multiple interconnected communities and may adopt norms that conflict with those of their home instances~\cite{zhang_troubleinparadise}. As the network grows, moderators increasingly face decisions about inter-instance relationships including federation, defederation, and content filtering.  
Understanding how moderation functions under these conditions is essential both for sustaining decentralized, volunteer-driven systems and for informing how self-governance could adapt in environments where communities are highly interconnected.

Community rules are a key component of self-governance. Rules formalize norms and guide moderation decisions such as curating content, developing new guidelines, and managing conflicts~\cite{seering2019moderator}.
As a record of accepted social practices, written rules align the understandings of members and moderators about expected behaviors~\cite{Koshy_usermodalignment}. Referencing rules can further reinforce shared norms ~\cite{fang2023shapingonlinedialogueexamining, Cai_moderationvisibility}. Educating about the rules is an effective moderation strategy to deter problematic users~\cite{cai2019effective} on Twitch, and increase rule compliance as well as positive participation on Reddit~\cite{bruckman_reddit_automoderator}. 
At the same time, rules do not fully capture how governance is enacted: in practice, enforcement may diverge from formal policy depending on context and discretion~\cite{matias2016civic}. 
Nevertheless, because rules are publicly visible, persistent, and comparable across communities, they offer a valuable lens for examining self-governance at scale---particularly in decentralized systems where other governance signals are often inaccessible or inconsistent.

Across platforms, larger, older, and more engaging communities tend to have more extensive rule sets and more moderation activities~\cite{frey_governing_2022, bhattacharya2024unveiling}. This pattern may reflect both selection effects---where communities with formalized rules are more resilient to abuse---and adaptation effects, where growth introduces new challenges, such as information overload and decreased interaction quality~\cite{nematzadeh2019information} that necessitate more formal governance~\cite{webber_fractal_2020}. 
In this work, we draw on Star and Ruhleder to conceptualize rules as governance artifacts that surface otherwise backgrounded coordination work under conditions of infrastructural strain, such as changes in community size \cite{star1994steps}. We examine how the characteristics of community rules vary across Mastodon instances of different sizes to understand how self-governance adapts under changing organizational conditions, while recognizing that rules represent only one facet of broader governance processes.

Prior research has documented how rules formalize with community growth on centralized platforms like Reddit~\cite{frey_governing_2022}, where platforms can invest in moderation infrastructure as communities scale. However, it remains unclear whether these patterns hold in decentralized, volunteer-driven systems where each community must independently address scaling challenges without centralized resources or coordination. This gap is significant: if scaling dynamics differ in federated systems, it would suggest that governance tools developed for centralized platforms may not transfer effectively; if they are similar, it would indicate fundamental constraints on self-governance that transcend platform architecture.

We systematically analyze rule variation across instances to examine how moderation practices scale in decentralized social media. We ask the following research questions:
\begin{itemize}[label={}]
    \item \textbf{Q1:} How do moderation priorities on Mastodon vary with community scale? 
    \item \textbf{Q2:} How is community scale associated with rule formalization, such as scope and lexical features?
    \item \textbf{Q3:} To what extent do rules reflect internal needs versus external federation interactions?
\end{itemize}

Our study builds on earlier work that categorized community rules on Mastodon~\cite{nicholson_mastodon_2023}, extending both the scope and analytical depth. 
We introduce a large-scale, multi-lingual dataset with 28,910 rules across 6,660 instances of size one to over two million. 
Unlike existing research that focuses only on popular, large, English-speaking Mastodon instances, our dataset allows us to analyze rules across the entire active ecosystem, including small communities that are often preferred by those seeking support, such as LGBT communities~\cite{zhang_troubleinparadise}. 
This rich dataset allows us to examine scaling behaviors that would be otherwise obscured in other smaller or instance-limited samples, enabling a comprehensive view growth dynamics and coordination on Mastodon.

Using this dataset, we make three contributions.
First, we show that governance priorities are remarkably consistent across instance sizes: rules addressing harassment, hate speech, and illegal content dominate regardless of scale, suggesting normative convergence even within a decentralized ecosystem.
Second, we demonstrate that scaling patterns observed in centralized platforms---where larger communities develop more extensive but less readable rules---persist in volunteer-driven, decentralized environments, even without monetary incentives or centralized support. This finding suggests fundamental constraints on self-governance that transcend platform architecture.
Third, we find that rule formalization is primarily associated with internal community size rather than external federation interactions, indicating that local scaling pressures outweigh network-level dynamics in shaping governance practices.
 
\section{Related work} \label{Related works}

Ample work has established that decentralization empowers community self-governance, that moderation strategies must adapt as communities scale, and that rules tend to formalize and proliferate as communities grow or encounter new challenges. Yet, how these dynamics unfold in federated, volunteer-moderated platforms like Mastodon remains under-explored. By analyzing Mastodon communities of varying sizes, our work connects these threads and fills a gap in understanding how rules evolve and formalize in decentralized contexts. In this section, we discuss previous literature and set the stage for our inquiry into Mastodon’s moderation.

\subsubsection*{Decentralization and Community Self-governance}

Elinor Ostrom demonstrated that communities can effectively self-regulate common-pool resources without centralized control, guided by design principles such as clearly defined boundaries, participatory rule-making, and nested governance~\cite{ostrom1992governing, ostrom2000collective}. Drawing on Ostrom, early research on online communities---such as Kollock and Smith’s work on ``virtual commons''---showed how communities manage shared spaces through collective norms and self-enforcement~\cite{kollock2011managing}. Herring further illustrated how early forums and chat groups developed their own codes of conduct and mechanisms for upholding norms~\cite{herring2004cmda}. In the context of Wikipedia, studies on decentralized governance also found that communities can create and enforce complex rule systems when granted autonomy~\cite{bruckman_decentralization_wikipedia}.

Decentralized social platforms such as Mastodon embody these self-governance ideals by design. Each instance employs local governance, with administrators tailoring rules to fit their community’s values, fostering diverse cultures and social dynamics~\cite{zhang_troubleinparadise, la_cava_understanding_2021, zignani_follow_2018, bono_exploration_2024, sabo_decentralized_2024, huang_decentralized_2024}. This heterogeneity makes decentralized ecosystems a great setting for examining ``governable spaces''---infrastructures that empower community agency and participatory norm enforcement~\cite{schneider2022governable, lampe_crowdsourcing_2014}.

Most governance research to date has focused on single-platform ecosystems like Reddit or Wikipedia or on policy decisions made by centralized corporate platforms~\cite{reddy_evolution_2023, Fiesler_Jiang_McCann_Frye_Brubaker_2018,gillespie_custodians_2018}. Scholars are now beginning to adapt governance frameworks to decentralized settings. For instance, the concept of multi-level governance---spanning platform-level protocols, community-level rules, and individual behavior---is increasingly applied to study the complex power dynamics in federated systems~\cite{jhaver_decentralize_platpower}.

\subsubsection*{Scaling of Moderation Strategies in Online Communities}

Institutional theory suggests that as groups grow in size and complexity, informal norms are increasingly codified into formal rules to reduce ambiguity and maintain stability \cite{scott_institutions_2013}. From a social identity perspective, larger communities also heighten the need for clear boundaries and behavioral expectations, making formalized rules an important mechanism for reinforcing shared identity \cite{ashforth_social_1989}. Empirical work by prominent scholars on online communities such as Preece and Seering has shown that the age of a community can influence governance: over time shared norms are created, and social capital accumulates, which in turn shapes formalization of rules \cite{PreeceMaloneyKrichmar2005, seering_reconsidering_2020}. Mahony and Ferraro further highlight a direct link between community age and increasing formality and stability of its rules \cite{OMahonyFerraro2007}. This body of work suggests that community size and age is a critical explanatory factor in understanding the formalization and development of community rules.

On many platforms, the burden of moderation falls to volunteers. Seering et al. highlight how Twitch and Reddit rely on volunteers as the first line of governance, reflecting a ``many communities, many local governors'' model within a broader system~\cite{seering_moderator_2019}. Klonick similarly describes volunteer moderators as part of the ``new governors'' of online speech, wherein private platforms and their user moderators collectively create a multi-layered governance structure~\cite{klonick2017new}.

As communities expand, the challenge of moderation grows and moderators should increasingly rely on a combination of social and technical strategies. Nakandala et al., showed that as Twitch channels grow in popularity, moderation becomes more difficult to scale effectively: objectifying and gendered content is common in high-visibility channels but rare in less popular ones, illustrating how the popularity (scale) of the channels can erode community norms and moderation system that works for much smaller channels~\cite{Nakandala_Ciampaglia_Su_Ahn_2017}. 
Lampe and Resnick's analysis of Slashdot's distributed moderation system provided early evidence that community-based rating and meta-moderation can support large-scale moderation~\cite{lampe2004slash}. Community rule formalization is another tool to enhance moderation transparency and communication, as echoed in moderation strategies across case studies of Twitch~\cite{cai2019effective} and Reddit~\cite{bruckman_reddit_automoderator}. Indeed, many large online forums have adopted templates for removal reasons or ``user education'' messages, institutionalizing this strategy~\cite{seering_reconsidering_2020}. 

Federated platforms like Mastodon rely even more on volunteer-driven models, where moderation is guided by local norms, inter-instance agreements, and optional tools~\cite{huang_decentralized_2024}. Unlike centralized platforms such as Facebook, X, and Reddit, scaling moderation on Mastodon is not simply a matter of increasing automation or expanding moderation teams. Instead, the community-led, volunteer-driven approach faces inherent cognitive and organizational constraints. As communities grow, higher post-to-moderator ratios can diminish moderation effectiveness, impacting user satisfaction and perceptions~\cite{weld2025perceptionsmoderatorslargescalemeasure}. Social cognitive theory suggests that managing relationships, including those online, is limited by cognitive capacity~\cite{dunbar_social_2012, dunbar_online_2016}. Consequently, as communities scale, informal rules often evolve into more structured governance to address increasing complexity~\cite{webber_fractal_2020}.
Our work builds on the ``community self-moderation'' research put forth by Seering et al.~\cite{seering_reconsidering_2020}, examining how Mastodon’s decentralized nature constrains moderator management in the absence of centralized oversight.

\subsubsection*{Evolution and Formalization of Community Rules}

 
The evolution of community rules is shaped not only by growth in size but by the ways communities interact with one another. As groups expand over time, mechanisms of reciprocity and reputation weaken \cite{Hodgson_2025, powers2016}, a pattern especially prevalent in online communities where anonymity is valued. Thus, while informal, trust-based systems may suffice in small groups, larger communities depend on increasing formalization, including designated roles, written rules, and structured moderation to sustain cohesion \cite{GalloYan2015}.
Smith’s study of MicroMUSE, a 1990s educational MUD, documents how governance shifted from openness to formal gatekeeping in response to disruptive growth~\cite{smith2002problems}. 
Kraut and Resnick frame this transition in terms of community life cycles, noting that documented policies and official roles become necessary as shared understanding wanes with scale~\cite{kraut2012building}. 
Previous research shows that interacting communities often observe what each do and adopt, borrow, adapt and/or discard rules accordingly \cite{MarchSchulzZhou2000}.
Taken together, prior research suggests that the size of a community and its degree of federation might play a critical role in shaping Mastodon's rules.

Herring and others show that governance evolves not only structurally but linguistically: informal norms enacted through discourse eventually give way to codified rule language as communities grow~\cite{herring2004cmda}. Related work has explored how rule types and linguistic structure must adapt for effectiveness at scale~\cite{frey_emergence_2019, frey_governing_2022, bulat_psychology_2024}, and that larger communities tend to adopt more standardized, systematic linguistic systems~\cite{raviv_larger_2019, zhu-jurgens-2021-structure}.
However, the relationship between scaling of moderation and participation is more nuanced. Stricter moderation can improve content quality but at the same time introduce barriers to entry that might deter participation and inclusivity~\cite{kraut_building_member_attach, kraut2012building, Kraut_commonIdentity}. 

These insights highlight the role of well-crafted rules that are clear, readable, and context-sensitive as one solution to bridge these challenges
Our study contributes to this line of work by empirically analyzing how Mastodon instance rules vary in topic coverage, length, formality, and readability as a function of community size and age, offering insight into the evolution of rule structures in decentralized, volunteer-moderated platforms.

\section{Data} 
\label{data}

Data for this study was collected in August 2024 using the instances.social API, which covers all Fediverse platforms. Starting from 15,571 candidate instances, we retained only Mastodon instances that returned at least one rule via the Mastodon API (see Appendix~\ref{appx:data_collection} for filtering details).
Three outlier instances were removed upon manual inspection (see Appendix~\ref{appx:data_collection}). We also excluded mastodon.social from our analysis as it is the only instance with over 10 million users and, as the ``default'' instance, it may deviate from patterns observed in organic instances. For these instances, the following metadata was collected: Instance ID, Instance Name, User Count, short Description, and status flags indicating whether the instance is active (see Appendix~\ref{appx:data_collection}). As a result, our sample contains 6,660 instances across the size spectrum with 28,910 rules. 

Although moderation is local, federation exposes users to external content, increasing moderation demands. As instances connect to more diverse communities, dissonant content becomes more likely, raising the burden of boundary decisions such as defederation or bans~\cite{zhang_troubleinparadise,bruckman_deplatform}. To capture the role of federation degree in shaping moderation complexity, we used the Mastodon API to collect the following information: federation degree (number of federating connections) and weekly activity data, including numbers of statuses, logins, and new accounts during the week of October 14--28, 2024.

\begin{figure}
\centering
\includegraphics[width=\textwidth]{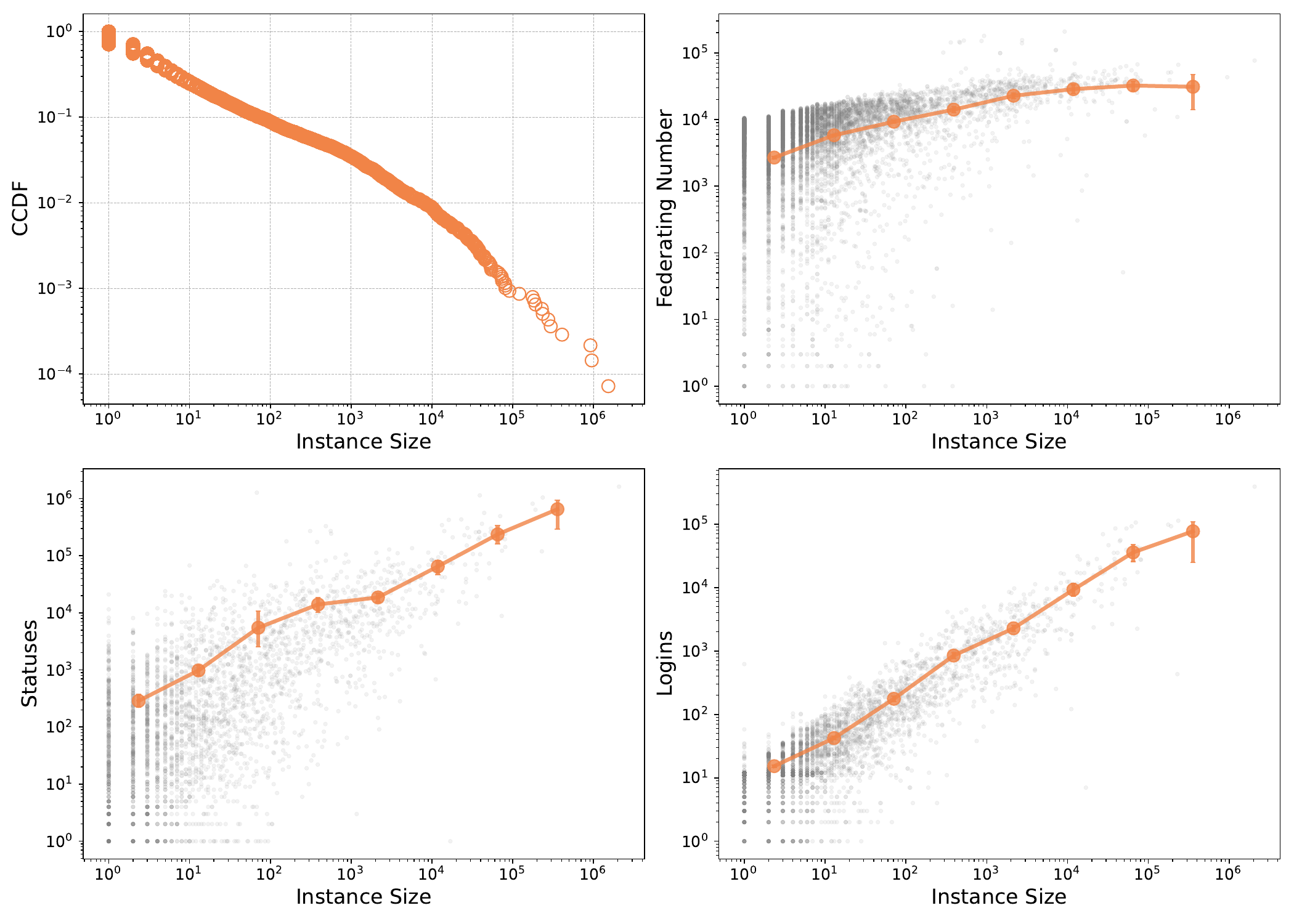}
\caption{Descriptive statistics of all Mastodon instances (a) Complementary cumulative distribution (CCDF) of instance size.
     Scatter plots of various activity metrics against instance size: (b) the number of federations, (c) the number of weekly statuses, and (d) the number of weekly logins. Orange dots represent the bootstrapped average and 95\% confidence interval (CI) for log bins of 5. 
    }  \label{fig:instance_vs_content}
\end{figure}

Figure \ref{fig:instance_vs_content} visualizes the descriptive statistics for the instances in our dataset, showing a heavy-tailed distribution of instance sizes. Activity levels, measured by the number of federations, the number of weekly statuses, and the number of weekly logins increase exponentially. 

Non-English rules account for a third of the dataset. We translated all non-English rules to English using Google Translate and validated translation quality via back-translation and readability comparison (see Appendix~\ref{appx:data_collection}).

\section{Methods} \label{Methods}

\subsection{Rule Categorization} \label{Rule Categorization}

We systematically analyze the topics of rules present across Mastodon instances to understand how communities formalize community norms. Each rule may be assigned multiple rule topics. For a clearer comparative analysis, we group the rule topics into four broader thematic categories---Action, Behavior, Content, and Distribution---based on their functional focus. 

For rule topic analysis of the rules, we randomly sample 1,100 unique rules and perform inductive coding. We find that 85\% of these rules align with the rule types defined in the framework developed  by~\cite{nicholson_mastodon_2023}, indicating substantial overlap between our emergent codes and their taxonomy. However, we modify this taxonomy to best fit our needs. This includes the removal of ``Prescriptive'' and ``Restrictive'' topic. In practice, these two topics were frequently conflated---many rules framed as restrictions also implicitly prescribed behavior (e.g., “Do not post hate speech” vs. “Respect others”), making it difficult to reliably distinguish between them. This ambiguity weakens their analytical utility and compromises inter-coder reliability. By collapsing these topics, we retain a sharper conceptual boundary across rule topics and reduce noise in our analysis of rule scope. We also eliminate certain topics that are too granular or ambiguous in definition and are covered by other topics such as ``Politics'' and ``Low-quality Content''. Using the annotated dataset from Nicholson et al.~\cite{nicholson_mastodon_2023}, which contains 690 rules from top Mastodon instances labeled with one or more topics, we develop a few-shot prompt to enable large language models (LLMs) to perform multi-label rule classification. We select few-shot prompting for its reproducibility and effectiveness in multi-label tasks~\cite{fewshot_multi_label}. We test the prompt on GPT-3.5-turbo and GPT-4o-mini, with GPT-4o-mini achieving the highest accuracy of 76\%, measured by comparing model-generated and manual annotations on a per-rule basis. Details about the prompts and the variants used for testing are mentioned in Appendix~\ref{appx:llm_clf}. We use this validated prompt to classify 28,910 rules in our full dataset.

To assess whether linguistic redundancy in rule phrasing could influence topic distributions, we conduct an additional robustness test wherein we deduplicate highly similar rules within each instance using a text similarity threshold using Jaccard Similarity. Repeating the topic analysis on this deduplicated dataset yields substantively similar results, indicating that our findings are not driven by stylistic variation or differences in how instances segment their rules (see Appendix~\ref{appx:deduplicate}). 

We adapt the Action, Behavior, Content, Distribution (ABCD) framework~\cite{abcd} to group the rule topics in Mastodon instances (Table.~\ref{table:abcd}). The framework, originally created for understanding platform policies for influence operations, clearly identifies the five following categories: ``Action'': operations through accounts, ``Behavior'': tactics used on the platform, ``Content'': messages, narratives and subject matter, ``Distribution'': reach of content and techniques including but not limited to duplicate accounts for re-posting and ``Effects'': actual consequences. We refine the framework to suit our analysis. First, we eliminate the ``Effects'' category as this requires analysis of moderation actions which is beyond the scope of the paper. Of the 19 rule topics we have, some mapping was straightforward. For instance, ``Hate Speech'', ``Illegal Content'' naturally belong in the Content category, while ``Spam'' and ``Reposting/ Crossposting'' belong in Distribution. For rule topics whose categories are ambiguous, we use a bigram co-occurrence matrix to decide how moderators might interpret the rules in practice. We use a simple tf-idf with bigrams and inspect the occurrence map to observe which rules coincide. For example, ``Copyright/ Piracy'' could be a Content rule type but it occurs with ``Advertising and commercialization'' which belongs in the Distribution category, thus ``Copyright/ Piracy'' is also categorized under the Distribution section.

\subsection{Rule Diversity} \label{Rule Diversity}
To measure rule diversity, which we define as the amount of variety among the categories of rules within a given instance, we use entropy. Entropy is a well-established measure of uncertainty or diversity in a distribution, making it well-suited to quantify how evenly rules are distributed across different categories. A higher entropy value indicates a more balanced and diverse rule set, while lower values suggest a concentration in fewer categories.
Let \( p_i \) denote the proportion of rules in category \( i \) (where \( i \in \{A, B, C, D\} \) for the four categories). The entropy \( H \) is calculated as:
\[
H_{\text{ABCD}} = - \sum_{\substack{i = \text{A}, \text{B}, \text{C}, \text{D}}}^{Q} p_i \log_2 p_i
\]

We map each rule to one of four categories: Action, Behavior, Content, and Distribution. Since we use a base-2 logarithm and there are four categories, the entropy ranges from 0 (no diversity; all rules fall into one category) to 2 (maximum diversity; rules are evenly distributed across all four categories). The number and variety of rule categories allude to the scope of rule formalization that instances of different sizes account for. 

\begin{table}[ht]
\centering
\caption{Categories of Mastodon rules}
\begin{tabular}{@{}p{0.11\textwidth} p{0.22\textwidth} p{0.54\textwidth} c@{}}
\toprule
\textbf{Category} & \textbf{Rule Topic} & \textbf{Example} & \textbf{\% Instances} \\
\midrule
Action & Automated tools & \textit{Bots need instances admin approval} & 2.24 \\
\midrule
\multirow{5}{*}{Behavior}
& Harassment & \textit{Do not engage in harassment of any kind.} & 11.58 \\  
& Doxxing & \textit{No harassment, dogpiling or doxxing of other users} & 5.54 \\ 
& Brigading/Dogpiling & \textit{No harassment, dogpiling or doxxing of other users} & 1.14 \\
& Impersonation & \textit{No impersonation of individuals, public figures, or organizations unless clearly marked as parody.} & 1.57 \\
& Trolling & \textit{ABSOLUTELY NO TROLLING! This instance has ZERO TOLERANCE for trolls!} & 1.44 \\
\midrule
\multirow{9}{*}{Content}
& Hate Speech & \textit{Hate speech is not tolerated. This includes but is not limited to racism, sexism, homophobia, transphobia, classism, ableism, etc.} & 18.03 \\
& NSFW & \textit{No explicit (NSFW) content without content warnings or sensitive media markers. Must not appear in avatars or headers.} & 8.31 \\
& Content Warnings & \textit{Use content warnings for text, images, video, and links appropriately.} & 7.09 \\
& Illegal Content & \textit{Abide by the law. No illegal content whatsoever.} & 9.99 \\
& Incitement of Violence & \textit{No incitement of violence or promotion of violent ideologies.} & 5.95 \\
& Mis/Disinformation /Conspiracy & \textit{No spreading of mis/disinformation or conspiracy content.} & 5.46 \\
& Images & \textit{No explicit sexual or violent images.} & 5.25 \\
& Off-topic/Topic Specific & \textit{Keep posts relevant to the instance’s topic.} & 1.72 \\
& Links \& Outside Content & \textit{No spammy or inappropriate external links.} & 1.54 \\
\midrule
\multirow{4}{*}{Distribution}
& Spam & \textit{Spam is not authorized or welcome.} & 6.24 \\
& Advertising \& Commercialization & \textit{Repeated or intrusive advertising is not accepted. Accounts may be silenced or removed.} & 4.1 \\
& Copyright/Piracy & \textit{You must obtain permission to quote, reproduce, or adapt any post.} & 2.11 \\
& Reposting/Crossposting & \textit{No reposting or crossposting of copyrighted material.} & 0.7 \\
\bottomrule
\end{tabular}
\label{table:abcd}
\end{table}

\subsection{Lexical Features} \label{Lexical Features}

A lexical feature is a characteristic of the words or vocabulary in a text, such as word frequency, diversity, length, or usage patterns, that reflects its linguistic and structural properties. Previous research suggests that lexical features can provide valuable insights into the dynamics and governance of communities. In this study, lexical features were analyzed to examine the linguistic characteristics of community rules and their evolution as community size increased. 
We use number of users as a proxy for community size, as it is readily available across instances and consistent with prior work on online community scaling~\cite{frey_governing_2022}. This choice aligns with theories of governance scaling, which emphasize coordination and rule-setting costs that grow with the size of the membership rather than solely with interaction frequency. While content production and visible activity are driven by participating users, governance and moderation must account for the entire membership, including inactive or “lurking” users, often in a proactive rather than purely reactive manner~\cite{lampe_motivations_2010}. Another meaningful proxy could be engagement volume which captures behavioral intensity but does not fully represent the population over which governance applies.

However, to ensure robustness, we repeat the following analysis using number of statuses (engagement) as a proxy for community size instead of user count (see Appendix~\ref{appx:alt_measure}).

\begin{itemize}
    \item \textit{Word Count}.
Prior studies \cite{raviv_larger_2019} indicate that word count tends to increase with community size. As communities grow, more words are used to express explicit expectations and describe a larger variety of challenges in hopes to provide participants with all required information about the community.

\item \textit{Rule Count}.
While rule count is a structural feature rather than a lexical feature, we include it in our analysis as number of rules can indicate scope of the rules. Further, the presence of more rules can lead to a richer or more complex lexicon. More rules can also indicate legal or technical vocabulary and rule count can be a proxy lexical measure. 

\item \textit{Type-Token Ratio (TTR)}
This metric is used to quantify the diversity of vocabulary within a text~\cite{johnson1944studies}. It is calculated by dividing the number of unique words (types) by the total number of words (tokens) in the text. A higher TTR indicates greater lexical variety, while a lower TTR suggests more repetitive or limited word usage. This metric is often employed to analyze language complexity or assess the lexical richness of written or spoken material. For this study, we focus on instances where TTR scores were meaningful by excluding rules containing fewer than seven words.

\item \textit{Readability Score}
The Flesch-Kincaid Reading Ease (FRE) score is a measure of assessing of a given text \cite{flesch_new_1948}. Here, we use the set of rules of an instance to analyze readability. We measure readability as it is a signal of comprehension. We hypothesize, the larger the instance is, the more comprehendible a rule should be as the rules serve as a guideline to a larger and more diverse set of people. 

\end{itemize}

\subsection{Regression} \label{Regression}

We investigate two independent variables that might affect rule formation, instance size and federation degree. The dependent variables are lexical and topical features: word count, rule count, TTR and readability score.
We regress these outcomes on log-transformed instance size and federation degree, using model specifications appropriate to each outcome type. Count outcomes, i.e., word count, rule count, topic count, are estimated with negative binomial regression. or TTR, a ratio outcome with values between 0 and 1, we use beta regression with a logit link. Readability (Flesch Reading Ease, FRE) is estimated via OLS with HC3 robust standard errors.

Due to the nature of the data, the sample size for each of these model varies. In particular, after removing instances with zero federation, we get a sample of N=$4,800$ (for word and rule count model). For topic count, we further exclude instances where no rules could be categorized by the LLM, leaving N=$3,621$.
Japanese-script instances are omitted from TTR and readability models due to incompatibility with space-tokenized lexical metrics.

A simple rule count may inflate differences across instances by treating semantically similar norms as multiple distinct rules, e.g., a single norm might be split into separate rules such as ``Respect others' identities in full'' and ``Respect others' right to engage or disengage in conversation'', or versus combined into one rule such as ``Respect others' identities and their right to engage or disengage in conversation.''). We also use topic count as an outcome to capture the number of unique rule topics of an instance rather than the raw number of rule items retrieved from the API. 

The age of an instance could be a potential confounder in the relationship between size, federation, and rule characteristics. To assess this, we replicate all models on the subset with available creation dates (N=$3,356$) and estimate a sensitivity model controlling for age (see Appendix~\ref{appx:regression}).

\section{Results} \label{Results}

\subsection{Topics} \label{Results_topics}

\begin{figure}
\centering
\includegraphics[width=0.8\textwidth]{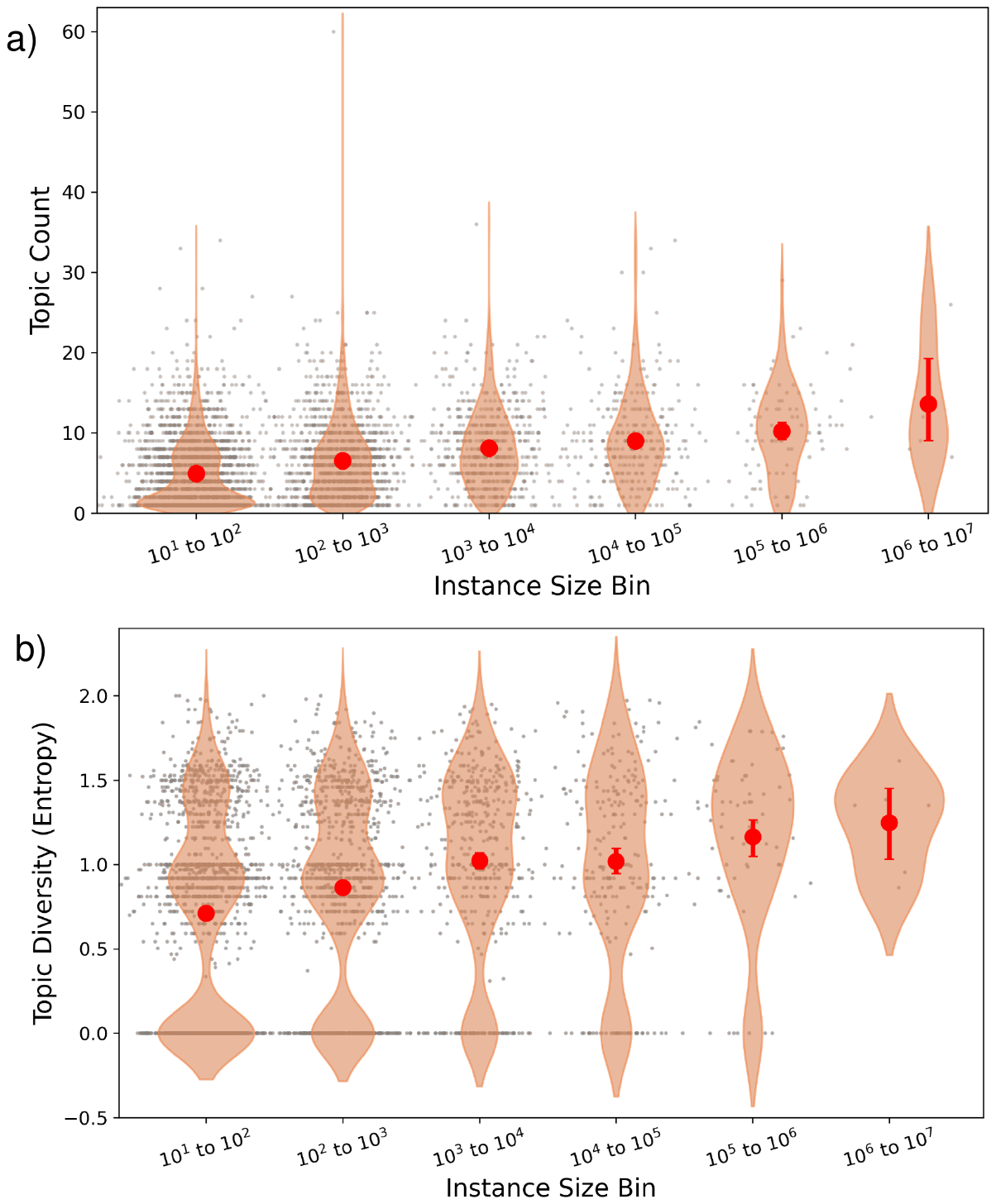}
\Description{Violin plot showing the distribution of number of topics per rule in instance bins.}
    \caption{Distribution of topic count, measured by number of unique rule topics per instance (top) and topic diversity, measured by entropy (bottom), across instance size bins.}
    \label{fig:violin_plot}
\end{figure}

\begin{figure}
\centering
\includegraphics[width=\textwidth]{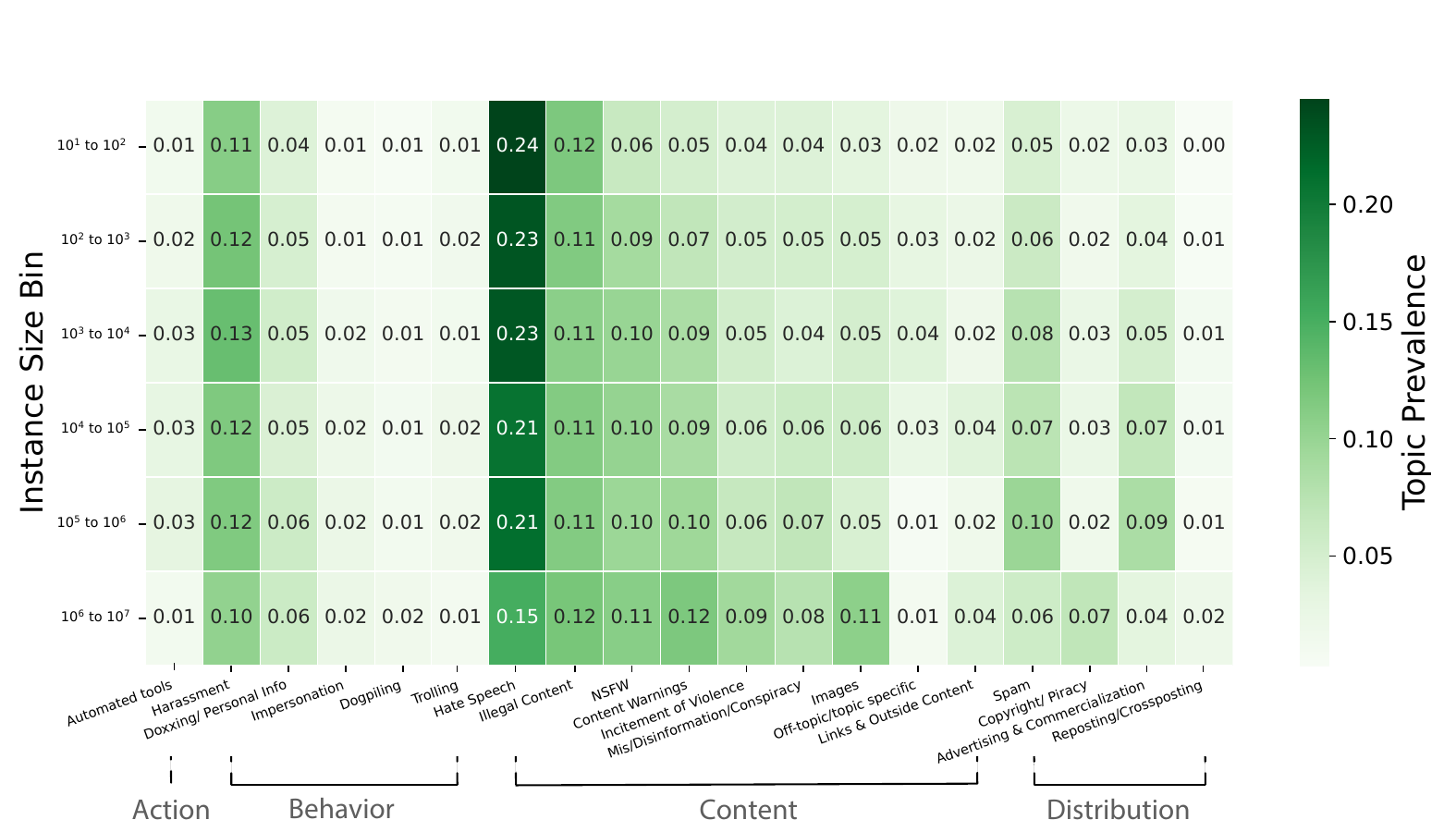}
\Description{}
\caption{Heat map of topic prevalence per instance, across instance size bins. Topics are grouped into one of the four categories: Action, Behavior, Content, and Distribution. Darker cells indicate higher prevalence of a given rule topic within an instance size bin.}
    \label{fig:topic-heatmap}
\end{figure}

Through rule classification, described in Section~\ref{Rule Categorization}, we found that ``Hate Speech'', ``Harassment'' and ``Illegal Content'' are a theme across the platform regardless of size, which is in line with previous research~\cite{reddy_evolution_2023, nicholson_mastodon_2023, Schaffner_communityguidelines}. The number of rules in an instance increases with size (Spearman's coefficient of $0.3$, p$<0.05$, Figure~\ref{fig:violin_plot}a). Our topic analysis (Fig.~\ref{fig:topic-heatmap}) reveals that smaller communities focus disproportionately on a few rule types, while larger instances maintain a balanced emphasis across many rule types. The results of our topic analysis trend remain consistent even after deduplication of linguistically similar rules in an instance (see Appendix~\ref{appx:deduplicate}). Community-topic focused analysis indicates that rule distributions do not differ substantially across community topics (see Appendix~\ref{appx:topical_focus}).

Our results of mapping rule types to the ABCD framework and measure of rule diversity (Fig.~\ref{fig:violin_plot}b) indicate that though some instances regardless of size have no diversity, on average larger instances tend to have greater diversity. This is supported by the Spearman's co-efficient measured between size and rule diversity of $0.21$ (p$<0.005$).

\begin{figure}
\centering
\includegraphics[width=\textwidth]{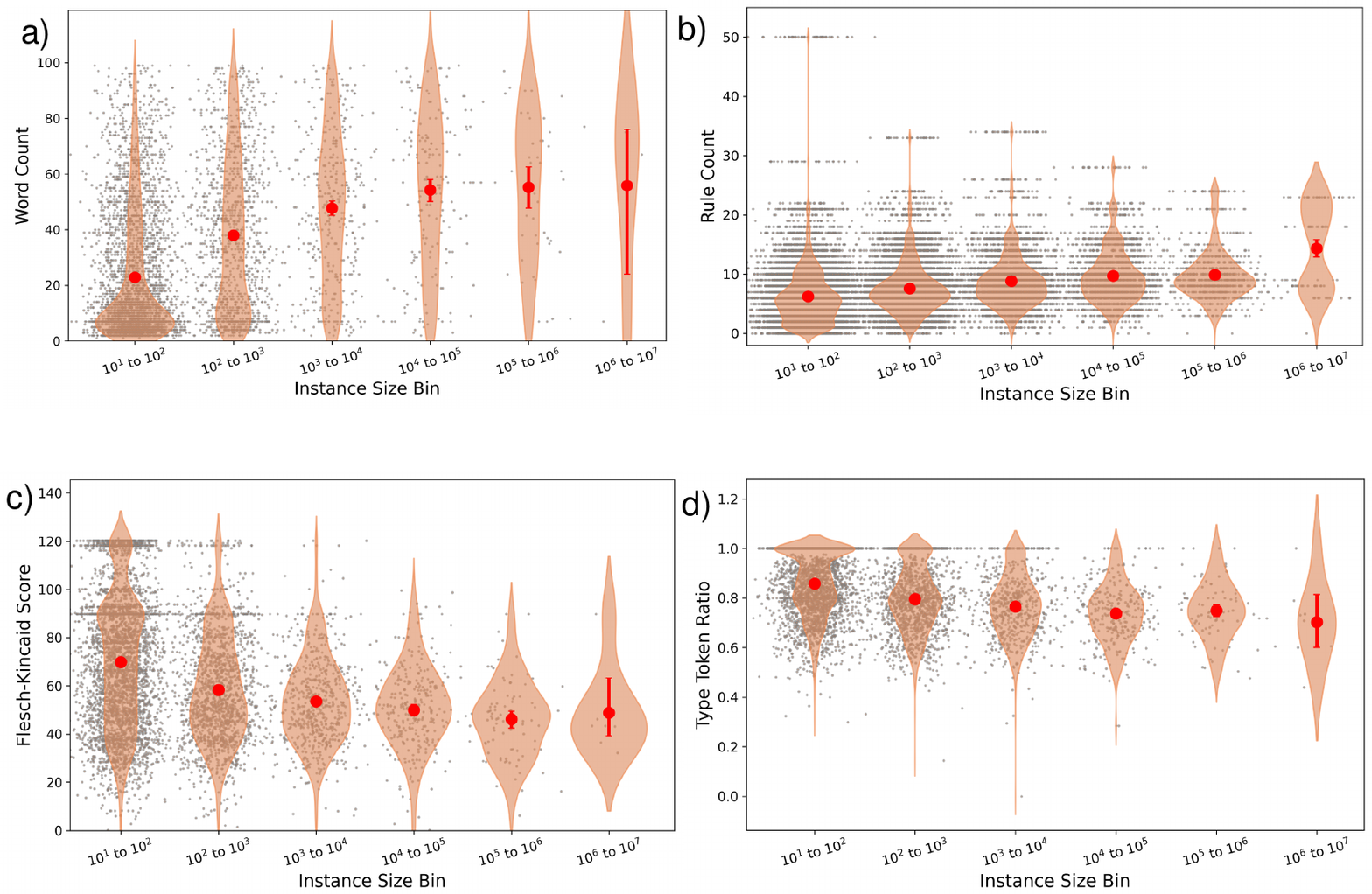}
\Description{}
\caption{Distribution of lexical features across instance sizes: (a) word count, (b) rule count, (c) readability of rule, measured by Flesch-Kinaid score, and (d) rule lexical diversity, measured by Type-Token ratio. Grey dots represent the values of features, while red dots and whiskers indicate the bootstrapped means and 95\% CI, respectively.}
    \label{fig:lexical-features}
\end{figure}

\subsection{Lexical Features} \label{Results_lexical_features}

The word count of the rules increases with instance size (Spearman's coefficient $\rho=0.40$, Fig.~\ref{fig:lexical-features}a). Rule count also increases as instances get larger ($\rho=0.33$, Fig.~\ref{fig:lexical-features}b), reflecting the need for more granular governance in larger communities. 
Somewhat counterintuively, larger communities tend to have lower rule readability ($\rho=-0.31$, Fig.~\ref{fig:lexical-features}c). Lexical diversity similarly decreases with instance size ($\rho=-0.34$, Fig.~\ref{fig:lexical-features}d), indicating greater vocabulary repetition in larger instances' rule texts. This likely reflects increasing standardization of rule language rather than simplification, as readability simultaneously decreases with scale. All correlations reported are statistically significant, $p<0.005$. We discuss potential explanations for these trends in Section~\ref{Discussion}.

Our analysis with engagement as a proxy of community size shows similar trends, where word and rule count increase as size increases, and readability and lexical diversity decrease with size (see Appendix~\ref{appx:alt_measure}). The correlations are consistently weaker under the engagement proxy, suggesting that membership, including inactive and lurking users, creates governance pressure beyond what behavioral activity alone predicts. This is consistent with a governance-burden interpretation in which moderators must formally address the all membership, including potential bad actors who may not yet have posted~\cite{lampe_motivations_2010}.

\subsection{Regression} \label{Results_Regression}

Table~\ref{table:nb_regression} shows the regression results of instance size and federation degree. Across models, we find that instance size is a consistent and statistically significant predictor for all rule formalization outcomes. Larger instances are associated with higher word count, rule count, and topic count ($\beta \ge 0.2$, $p<0.001$), as well as lower TTR ($\beta=-0.47$, $p<0.001$), and readability scores ($\beta=-8.14$, $p<0.001$), indicating longer and more topically diverse but lexically less varied and less readable rule descriptions.

\input{CSCW/table_primary}

For negative binomial regression estimating $\text{log}(E[Y]) \approx \beta X$, the coefficients represent log incidencerate ratios; we interpret the results using the exponentiated coefficients, Incidence Rate Ratios, $IRR = e^\beta$. Each tenfold increase in users is associated with 53\% more words in rule text (IRR=1.53), 34\% more rules (IRR=1.34), and 23\% more unique rule topics (IRR=1.23). 
For TTR, estimated via beta regression with a logit link, coefficients are on the log-odds scale and not directly interpretable. We therefore compute average marginal effects (AME), defined as $\frac{1}{N}\sum_i \beta \cdot \mu_i (1-\mu_i)$, which average the marginal effect $\partial \mu_i / \partial x$ over all observations. For log user count, this yields AME $\approx -0.06$, indicating that a tenfold increase in users is associated with a 7\% relative reduction at the sample mean (TTR $\approx 0.83$).
For the OLS model, coefficients are direct changes in FRE score on a 0–100 scale, where higher indicates more readable rules. 
Larger instances also show meaningfully lower rule readability, i.e., $-8.1$ FRE points per tenfold increase in users; moving from ``standard'' to ``fairly difficult'' on the Flesch-Kincaid scale.

Federation degree shows a smaller and less consistent effect, only significantly associated with rule count ($\beta=0.08$, $p<0.001$) and TTR ($\beta=-0.11$, $p<0.001$). 
Each tenfold increase in federation connections is associated with 9\% more rules (IRR=1.09, $p<0.001$).
Federation's effect on TTR ($-0.01$ units at the mean) is statistically significant but negligibly small in practice.
This suggests that instances with more cross-instance exposure have modestly broader the scope of governance, but  no difference in linguistic characters. 

The sensitivity analysis including community age as a predictor, yields substantively similar conclusions (Appendix~\ref{appx:regression}).

\subsection{Robustness}

Previous research indicates that communities centered on similar topics often develop similar norm sets \cite{reddy_evolution_2023}. At the same time, Mastodon has also served as a major destination for users migrating from Twitter/X, particularly during October 2022 \cite{zia_flocking_2023, jeong2024exploringplatformmigrationpatterns}, which could also have influenced instance-level moderation rules.
These dynamics raise the possibility that both topical focus and migration dynamics may confound the observed patterns in rule formation.

We therefore examine each factor in turn. First, to assess the effect of topical focus, we manually categorize the short descriptions of instances and compute the rule similarities across instance description topics (see Appendix~\ref{appx:topical_focus}). We find that more than 80\% of the topics has high cosine similarity of above 0.92, indicating no significant difference between the rule types of different topics (Fig.\ref{fig:topic-category-ccdf}). 

Second, to evaluate the effect of the migration on instance rule sets, we used Wayback Machine\footnote{\url{web.archive.org}} to recover the rules of instances at three time points: October 2022 (Twitter/X acquisition), January 2023 (shortly after the migration surge), and January 2024 (one year later). Only a small set of 60 instances were recoverable and available for longitudinal comparison. We categorize and compare the rule sets at different time points for these instances using Jaccard similarity (see Appendix~\ref{appx:wayback}). Instances are considered to have changed if the Jaccard similarity of their rule sets between two time points is below 0.9. Yet with this relatively lax threshold, we find that only 19\% of instances changed their rule sets between 2022--2023, while only 10\% changed between 2023--2024.

Overall, we find limited evidence that the Twitter takeover systematically reshaped written rule sets (see Appendix~\ref{appx:wayback}). However, our conclusions remain limited by the small number of recoverable instances: only 60 instances had recoverable snapshots across all three time points, of which 11 showed rule changes between 2022 and 2023. The absence of detectable change in this small and likely non-representative sample as null effects for the broader ecosystem should be interpreted with caution.
Nonetheless, these analyses suggest that neither topical focus nor migration dynamics explain the variation in rule formalization documented in our main analysis, supporting the robustness of our findings of the primary explanatory factors, community size and federation.

\section{Discussion} 
\label{Discussion}

This study examines self-governance on Mastodon, focusing on the relationship between community size and rule complexity across a wide range of Mastodon instances. 


First, we find that across instances of all sizes, rules consistently emphasize preventing harassment, hate speech and illegal content. This aligns with previous research of the most popular Mastodon instances, showing that the platform and its moderators prioritize creating safe spaces free from harassment, hate speech, and other common online harms~\cite{nicholson_mastodon_2023}. Notably, the relative emphasis on these topics remains invariant across community sizes, suggesting a broadly shared normative commitment to harm reduction. Despite operating independently from one another, most Mastodon administrators appear to align with broad legal requirements and share normative priorities, such as anti-harassment, with mainstream platforms~\cite{Schaffner_communityguidelines}. This consistency points to a form of normative convergence within a decentralized ecosystem.


Second, we observe that rule formalization is strongly associated with instances size. Larger communities tend to adopt broader and more diverse rule sets, reflected in both topical scope (the number and the diversity of topics), and in rule volume (word count and rule count).
This pattern suggests an expansion of governance concerns as communities grow, consistent with prior work on scaling in online communities~\cite{frey_governing_2022, bhattacharya2024unveiling}. 

Linguistic analysis reveals a more complex picture: although rules in larger communities tend to have less lexical diversity---indicating greater standardization--- this does not translate into improved readability.
One plausible interpretation is that larger instances use more specialized or formal language to reduce ambiguity and to address a broader range of behaviors. In contrast, smaller communities may rely on more informal language that is easier to read but less precise.

Written rules, a core governance tool, become more formalized as communities grow, suggesting that scale remains a key challenge for moderation. While decentralization distributes governance across instances, each instance continues to face internal scaling pressures as its local community expands. Our regression results reinforce this: across a broad sample of 4,801 instances, community size is the strongest and most consistent predictor of rule formalization, with federation degree playing a secondary and outcome-specific role. This pattern holds in the age-controlled sensitivity model (N\,=\,3,356), confirming that the size effect is not an artifact of compositional differences between the two samples. These findings suggest that, despite inter-instance connections, governance practices are primarily associated with internal community needs and local interactions.

These results point to a broader insight: decentralization changes \textit{where} governance decisions are made, but not the fundamental relationship between community size and governance complexity. The persistence of similar scaling patterns across both centralized platforms (like Reddit) and decentralized, volunteer-driven systems (like Mastodon) suggests that the challenges of governing larger communities may be difficult to escape through platform architecture alone.

\subsection*{Design Implications for Moderation}

Taken together, these findings highlight an important design trade-off between precision and accessibility in community rules. More formal and detailed rules may support consistent enforcement and reduce ambiguity for moderators, but reduced readability can create barriers for users, particularly newcomers, and undermine rule awareness or voluntary compliance~\cite{kraut2012building}. Rules that are difficult to understand can shift interpretive power toward moderators, increasing discretionary enforcement and potentially affecting perceptions of fairness and legitimacy. 
In federated systems like Mastodon, where governance is local and compliance often relies on shared norms rather than formal sanctions, maintaining readable and inclusive rules is especially important as instances grow. 
Our results point to several concrete design opportunities for federated platforms. 

\textbf{Size-calibrated rule revisions.}
Rule formalization scales monotonically with instance size across all measures: scope, word count, topical diversity, and lexical complexity. Yet instance administrators do not currently receive any infrastructure anticipating this trajectory. 
Federated platforms could offer tiered rule templates calibrated to community size, providing a minimal high-readability template for new instances and detect when an instance crosses a size threshold (e.g., 100 or 1,000 users) and prompt administrators with structured templates for expanding their rule set or recruiting additional moderators. This would reduce the cold-start burden on administrators and promote governance continuity as communities expand~\cite{kraut2012building, seering_reconsidering_2020}.

\textbf{Readability tooling at the authoring stage.}
We find that larger instances produce rules with substantially lower readability scores ($\rho = -0.31$), despite serving larger and more linguistically diverse user populations where accessibility is most critical. To address this gap between complexity and accessibility, platforms could provide tools that show real-time readability feedback, analogous to spelling checkers, could nudge administrators toward plain-language rule formulation as complexity grows. The value of readable rules in supporting compliance and norm internalization is well established~\cite{cai2019effective, bruckman_reddit_automoderator}, and automated tooling could extend this benefit to contexts where administrators lack editorial support.

\textbf{Highlighting governance divergence across federated boundaries.}
The dominant rule categories---Hate Speech, Harassment, Illegal Content---are consistent across all instance sizes, representing a shared normative floor. Federation protocols currently offer limited visibility into \emph{above-floor} governance differences, which our data show do vary systematically with size and topical focus. Tools that surface these differences during federation negotiation---showing, for instance, that a prospective federation partner handles NSFW content or misinformation differently---could help administrators anticipate norm conflicts before they generate moderation load~\cite{jhaver_decentralize_platpower, zhang_troubleinparadise}.

\textbf{Resource allocation.}
Community size, not federation degree, is the dominant predictor of rule formalization across all outcomes. This finding suggests that platform support for governance would have greater payoffs when targeted at growing instances than at inter-instance coordination. This support  includes resources such as administrator training and moderation tooling.


\section{Limitations and Future Work} 

This study has several limitations. First, the analyses relies on a largely static snapshot of Mastodon. A longitudinal analysis examining how rules evolve as instances grow could offer deeper insights. 
Second, while our regression analyses identify strong associations between community size and rule formalization, they do not establish causality. Future work combining longitudinal data, qualitative methods, or direct observations of moderation practices could help unpack the mechanisms underlying these relationships.
Finally, our topic annotation relies on an LLM using a manually selected prompt. Although we carefully evaluated prompt performance, multi-label topic classification remains challenging and may introduce noise.

Our analysis of rule sets from various timestamps indicates that despite Mastodon being a key location for Twitter migration, the rule sets were not significantly affected. We find that many instances were created after the Twitter takeover, likely reflecting Mastodon’s increased visibility and adoption. Future work could examine the impacts of the Twitter/X migration to Mastodon more closely by combining longitudinal analyses of rule evolution with behavioral data, such as moderation actions or user participation patterns, to better understand how large-scale migration events influence moderation practices beyond formal rule articulation.


A second important direction for future research concerns the readability and accessibility of community rules. Further work could investigate how different design choices in rule presentation---such as summaries, examples, or layered disclosure---affect user understanding, perceptions of legitimacy, and moderation outcomes.

Finally, although some scholars advocate for network-level approaches to moderating federated systems~\cite{defederation2025}, our results indicate that rule formalization is most strongly associated with local community size rather than external federation interactions. Prior qualitative research, including interviews with Mastodon administrators, highlights tensions that arise when instances with differing norms interact~\cite{zhang_troubleinparadise, huang_decentralized_2024}. Future research should further explore how local autonomy and cross-instance coordination are negotiated over time, particularly as federated platforms continue to scale and diversify. Understanding this balance is critical for advancing theories of decentralized governance and for informing the design of tools that support coordination without undermining community self-governance.

\bibliographystyle{ACM-Reference-Format}
\bibliography{CSCW/refs}

\appendix \label{appendix}

\section{Regression: Age-Controlled Sensitivity Analysis}
\label{appx:regression}

Table~\ref{table:robust_regression} replicates the primary regression (Table~\ref{table:nb_regression}) on the subset of instances for which creation dates are recoverable via the Mastodon API (N=$3,356$ for count models; N=$3,059$ for TTR after Japanese exclusion). Community age is included as an additional predictor. Coefficients for instance size and federation degree are directionally consistent with the primary model. Age is negatively associated with rule count, word count, and topic count ($\beta=-0.02$ to $-0.04$, $p<0.001$), consistent with the interpretation that older communities maintain more concise rule sets, possibly reflecting rule maturation over time. Age is positively associated with lexical diversity ( $\beta=0.05$,$p<0.001$) and readability ($\beta=0.87$,$p<0.001$), suggesting that governance language becomes more varied and accessible as communities age.

\input{CSCW/table}

\section{Data Collection and Preprocessing}
\label{appx:data_collection}

The instances.social API provides instance information across all Fediverse platforms. We retrieved a list of all possible instances without any filters or parameters. This list is a reliable source for active instances in that it provides a  by default excludes instances that are ``dead'' instances, those that have been down for more than two weeks. Our collection yielded us a total of 15,571 instances of all languages and topics.

To ensure our dataset contained only Mastodon instances, we queried each instance URI through the Mastodon API to collect its rules. Of the 15,571 instances, 8,910 returned empty sets---either because they were not Mastodon instances, had no rules, or stored rules in locations not accessible through the API endpoint. Another 953 instances failed to retrieve rules due to unknown errors or server/client issues (e.g., 404, 5xx), which is a challenge also faced by prior research that collects data in the Fediverse \cite{anaobi_will_2023, tosch_privacy_2024}. 

For each instance, we collected the following metadata as defined by the instances.social API: Instance ID (unique ID number provided), Instance Name (unique name of the instance), User Count (number of users currently registered with the instance), Description (provides information about what the instance converses about), Dead (tells us if the instance is still up and running), Reg (if the instance is still allowing users to register). The last two fields are included as filters to ensure that all Mastodon instances collected are active and contributing to a valuable selection of instances.

Prior to the rest of the analysis, we removed outlier instances upon manual inspection: libra.net, verna.social, social.outsourcemath.com. The first two instances were legal communities with more than 150 rules, and the last had 500,000 federations but only one user, and was likely to be run by bots. 

Prior to our analysis of the rules, we conducted standard preprocessing such as making all the rules lower case and removing HTML entities, links and mentions. Upon manual inspection of a random set of rules with links, we found that the links lead to information about privacy settings of the instance, or links unrelated to written norms of the instance. Since our research questions are focused on the topic of the rule, the HTML entities and links are not relevant. 

\subsection*{Translation Quality}

As non-English rules account for a third of the rules in our dataset, we translated all non-English rules using Google Translate. We assess translation quality through two checks. First, we conducted back-translation and on a random sample of 120 rules covering a diverse range of languages and scripts. Each translated rule was re-translated back to the original language using Google Translate, and assessed on multiple aspects using GPT-4o-mini. We queried three questions from the model: 1) how much information loss is present? 2) how much contextual loss is present?, 3) Is there a significant difference in tone? GPT-4o-mini provides answer on a 3-point qualitative scale (``High'', ``Low'', ``Moderate''). In most cases, differences were related to tone. For example, the French phrase ``être tenus responsables'' (to be held accountable) was softened in translation to ``être responsables'' (to be responsible), implying a subtle but not substantive shift in assertiveness. Given the cultural and platform-specific nuances of tone, and the limited reliability of current models in capturing it, we chose not to include tone as a formal part of our evaluation.

Second, we evaluated the readability of the translated rules by comparing Flesch-Kincaid Reading Ease (FRE) scores between the original non-English rules and their English translations. FRE provides a standardized metric for assessing textual complexity and helps us determine whether the translation alters the intended reading difficulty of the rule. We expected a faithful translation to yield a small mean difference in FRE scores with low variance. We calculated the absolute difference in FRE scores for each rule and normalized by the maximum FRE score (100), resulting in values on a -1 to 1 scale. The normalized mean difference was 0 (SD=$0.13$), indicating that the translated texts maintain the structural complexity of the original rules. However, FRE can only be extended to a select number of European languages, limiting the generalization of the translation. To overcome this, we observed the lexical trends, namely word count and rule count, in the original rule text and the translated text. We found that the trends were similar, and thus conclude that the translation is robust for the purpose of this study. 


\section{Engagement as Alternative Proxy of Community Size}
\label{appx:alt_measure}

\begin{figure}
\centering
\includegraphics[width=\textwidth]{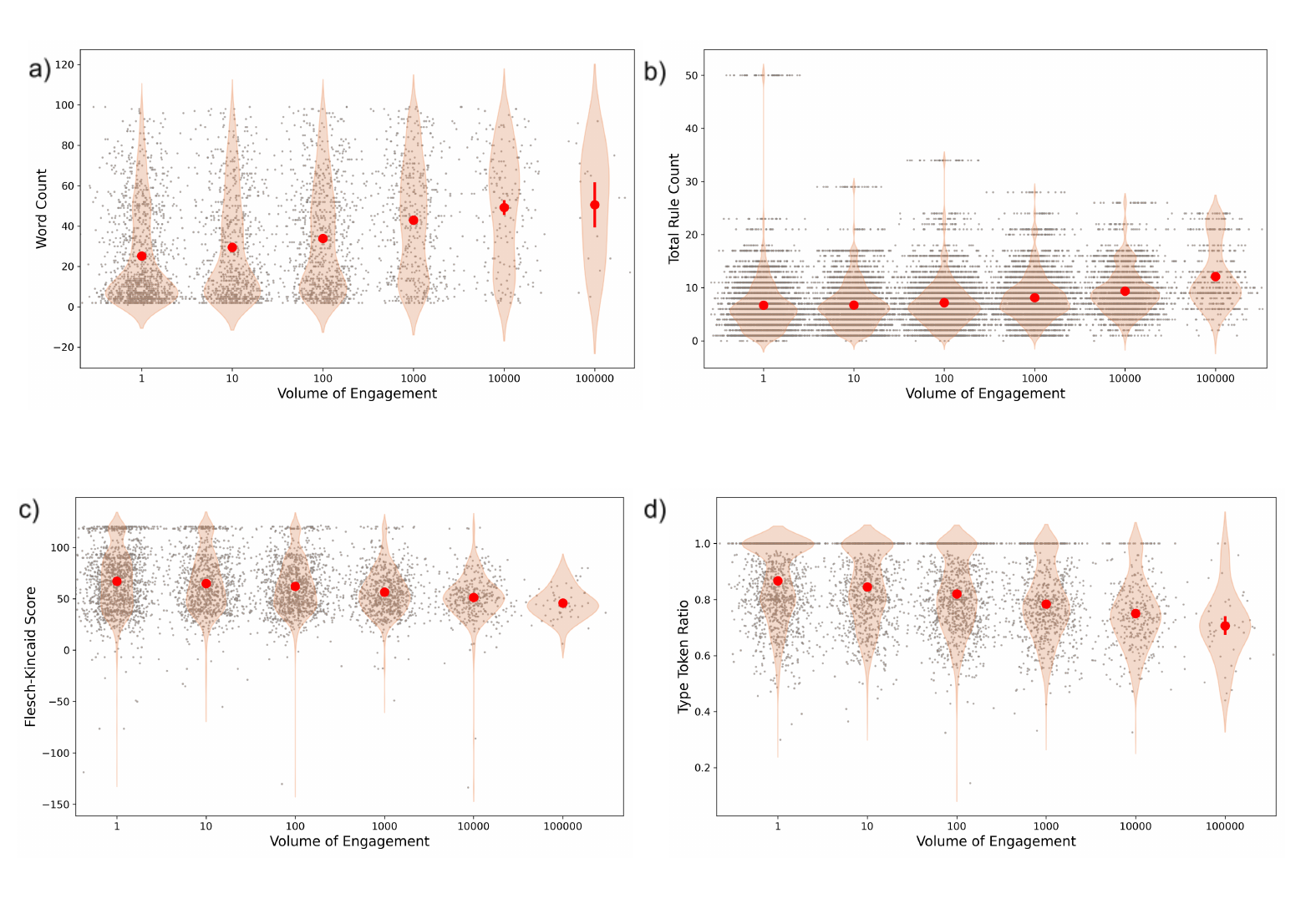}
\Description{}
\caption{Distribution of lexical features across engagement levels. (a) Rule count. (b) Word count. (c) Readability of rule, measured by Flesch-Kinaid score (d) Rule lexical diversity, measured by Type-Token ratio. Grey dots represent the values of features, while red dots and whiskers indicate the bootstrapped means and 95\% CI, respectively.}
    \label{fig:engagement-lexical-features}
\end{figure}

We investigate the robustness of user count as a proxy for instance size by replicating the analysis in Section~\nameref{Results_lexical_features}, replacing user count with engagement level. Engagement level is measured as the number of statuses posted per week within each instance. We collected these data over four weeks surrounding the data collection period and aggregated them accordingly.
The relationships between community size and lexical features are similar when using engagement level as a proxy. As engagement level increases, both word count (Fig.~\ref{fig:engagement-lexical-features}a, Spearman's coefficient is $0.27$ (p$<0.005$) and rule count increases (Fig.~\ref{fig:engagement-lexical-features}b, Spearman's coefficient is $0.26$ (p$<0.005$). As shown in Fig.~\ref{fig:engagement-lexical-features}c--d, lexical diversity decreases in more active instances (Spearman's $\rho = -0.28$), as does readability ($\rho = -0.17$), with both correlations statistically significant (p$<0.005$). This is consistent with our results using instance size.

\section{The Impact of Rule Redundancy}
\label{appx:deduplicate}

In a given instance, multiple rules may articulate the same underlying norm using slightly different language. For example, instances may treating semantically similar norms as multiple distinct rules (e.g., a single norm might be split into separate rules such as ``Respect others' identities in full'' and ``Respect others' right to engage or disengage in conversation'', or versus combined into one rule such as ``Respect others' identities and their right to engage or disengage in conversation.''). Although the normative content is substantively similar, differences in linguistic granularity can inflate the number of rules retrieved via the API. Such variation could inflate our topical analysis. 

To ensure that our results are not driven by superficial differences in language use or rule segmentation, we perform a deduplication procedure as a robustness check. Specifically, we first randomly sample pairs of rules from the entire dataset to establish a baseline distribution of semantic similarity. This gave us a threshold of 0.25. We then compare pairs of rules within each instance using Jaccard similarity and remove one rule from a pair if their similarity exceeds the threshold, treating highly similar rules as redundant expressions of the same norm. This procedure reduces the influence of stylistic or organizational differences in how instances articulate rules, while preserving substantively distinct norms.

We rerun our topic analysis on this deduplicated dataset and find that with increase in size, the emphasis on various topics become more equal. In comparison to smaller instances where an disproportional focus on few topics is present.

Additionally, instance size and topic count have a Spearmann correlation value of 0.324 (p<<0.05). This indicates that the overall trends do not change substantially. This suggests that our findings are not inflated by differences in linguistic granularity or redundant rule formulations, and that the observed patterns reflect substantive normative content rather than artifacts of rule phrasing or enumeration.

\begin{figure}
\centering
\includegraphics[width=\textwidth]{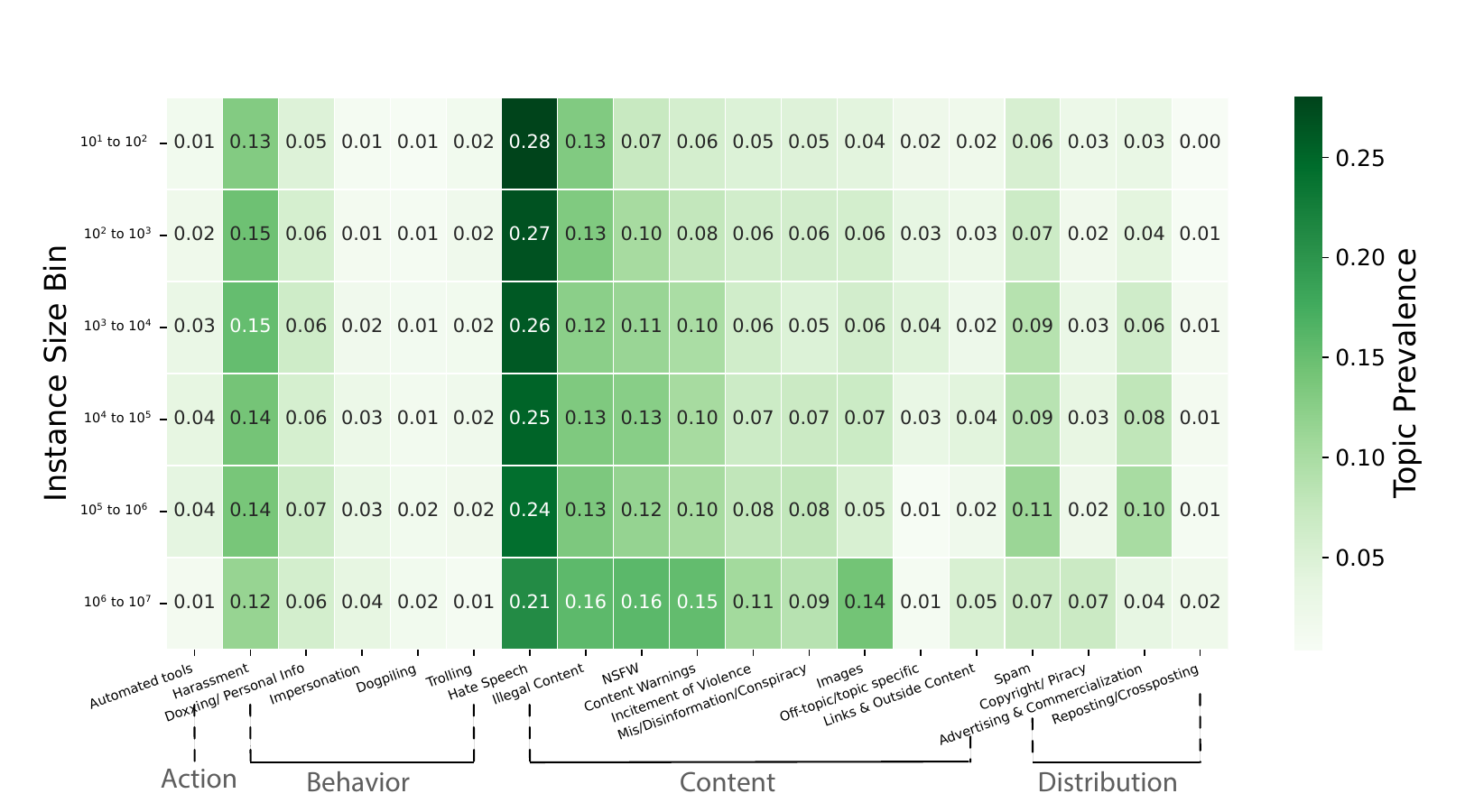}
\Description{}
\caption{Heat map of topic prevalence per instance, across instance size bins. Topics are deduplicated using a global Jaccard similarity threshold. Topics are grouped into one of the four categories: Action, Behavior, Content, and Distribution.}
    \label{fig:deduplicated-topic-heatmap}
\end{figure}

\section{LLM Classification Procedure}
\label{appx:llm_clf}

Our Rule Categorization, as described in Section~\ref{Rule Categorization}, uses a few-shot prompt (n=12) to categorize the 28,000 rules into categories adopted from~\cite{nicholson_mastodon_2023}. Here, we present the prompt and additional experimental and variant details. 

\begin{figure}[ht]
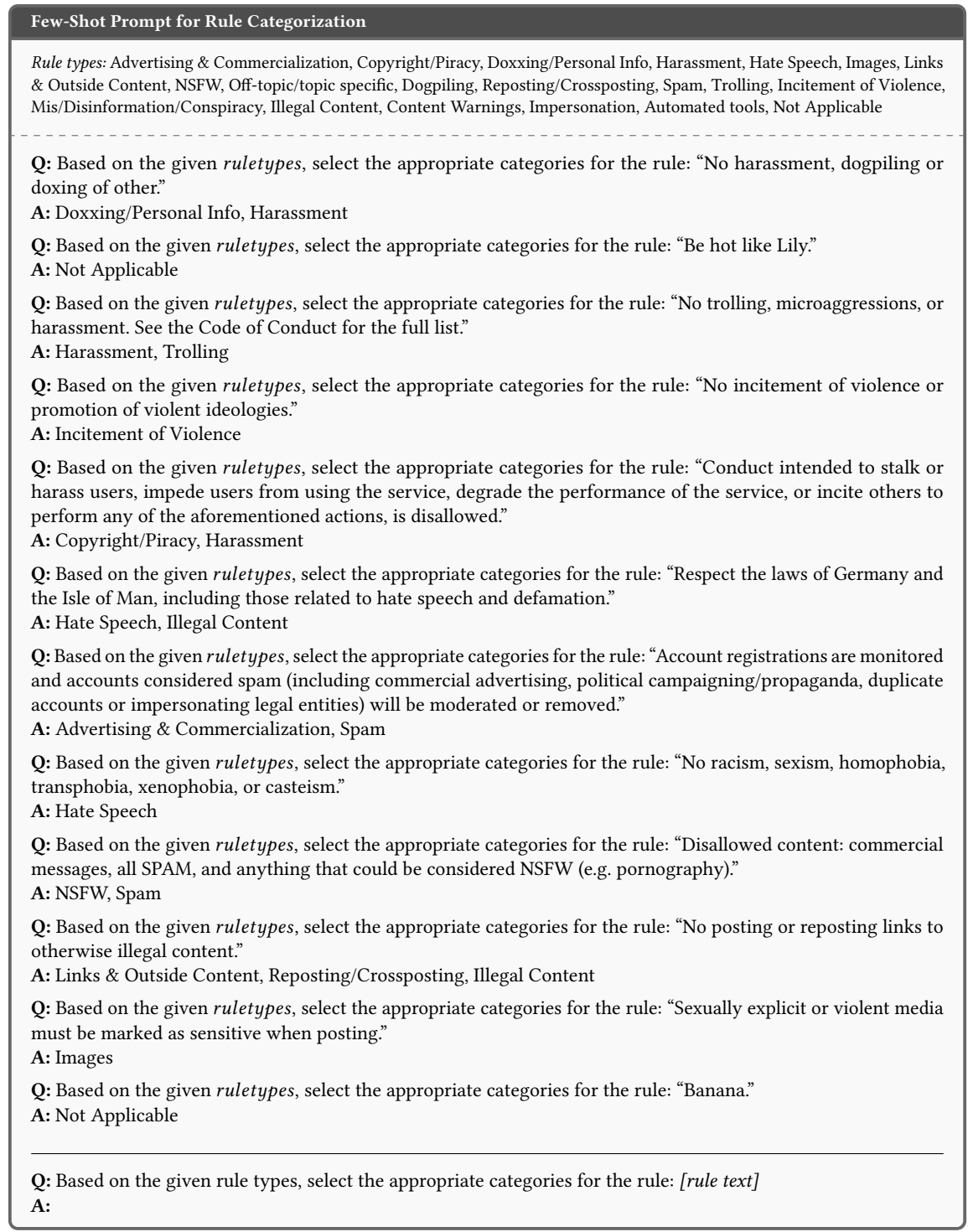

\centering
\begin{tcolorbox}[
  title=Few-Shot Prompt for Rule Categorization,
  colback=gray!5!white,
  colframe=black!60,
  coltitle=white,
  colbacktitle=black!70,
  fonttitle=\bfseries\small,
  left=6pt, right=6pt, top=4pt, bottom=4pt
]
\small
\textit{Rule types:} Advertising \& Commercialization, Copyright/Piracy, Doxxing/Personal Info,
Harassment, Hate Speech, Images, Links \& Outside Content, NSFW, Off-topic/topic specific,
Dogpiling, Reposting/Crossposting, Spam, Trolling, Incitement of Violence,
Mis/Disinformation/Conspiracy, Illegal Content, Content Warnings, Impersonation, Automated tools,
Not Applicable

\tcblower

\textbf{Q:} Based on the given ${rule types}$, select the appropriate categories for the rule: ``No harassment, dogpiling or doxing of other.''\\
\textbf{A:} Doxxing/Personal Info, Harassment

\medskip
\textbf{Q:} Based on the given ${rule types}$, select the appropriate categories for the rule: ``Be hot like Lily.''\\
\textbf{A:} Not Applicable

\medskip
\textbf{Q:} Based on the given ${rule types}$, select the appropriate categories for the rule: ``No trolling, microaggressions, or harassment. See the Code of Conduct for the full list.''\\
\textbf{A:} Harassment, Trolling

\medskip
\textbf{Q:} Based on the given ${rule types}$, select the appropriate categories for the rule: ``No incitement of violence or promotion of violent ideologies.''\\
\textbf{A:} Incitement of Violence

\medskip
\textbf{Q:} Based on the given ${rule types}$, select the appropriate categories for the rule: ``Conduct intended to stalk or harass users, impede users from using the service, degrade the performance of the service, or incite others to perform any of the aforementioned actions, is disallowed.''\\
\textbf{A:} Copyright/Piracy, Harassment

\medskip
\textbf{Q:} Based on the given ${rule types}$, select the appropriate categories for the rule: ``Respect the laws of Germany and the Isle of Man, including those related to hate speech and defamation.''\\
\textbf{A:} Hate Speech, Illegal Content

\medskip
\textbf{Q:} Based on the given ${rule types}$, select the appropriate categories for the rule: ``Account registrations are monitored and accounts considered spam (including commercial advertising, political campaigning/propaganda, duplicate accounts or impersonating legal entities) will be moderated or removed.''\\
\textbf{A:} Advertising \& Commercialization, Spam

\medskip
\textbf{Q:} Based on the given ${rule types}$, select the appropriate categories for the rule: ``No racism, sexism, homophobia, transphobia, xenophobia, or casteism.''\\
\textbf{A:} Hate Speech

\medskip
\textbf{Q:} Based on the given ${rule types}$, select the appropriate categories for the rule: ``Disallowed content: commercial messages, all SPAM, and anything that could be considered NSFW (e.g.\ pornography).''\\
\textbf{A:} NSFW, Spam

\medskip
\textbf{Q:} Based on the given ${rule types}$, select the appropriate categories for the rule: ``No posting or reposting links to otherwise illegal content.''\\
\textbf{A:} Links \& Outside Content, Reposting/Crossposting, Illegal Content

\medskip
\textbf{Q:} Based on the given ${rule types}$, select the appropriate categories for the rule: ``Sexually explicit or violent media must be marked as sensitive when posting.''\\
\textbf{A:} Images

\medskip
\textbf{Q:} Based on the given ${rule types}$, select the appropriate categories for the rule: ``Banana.''\\
\textbf{A:} Not Applicable

\medskip\noindent\rule{\linewidth}{0.4pt}\medskip

\textbf{Q:} Based on the given rule types, select the appropriate categories for the rule: \textit{[rule text]}\\
\textbf{A:}

\end{tcolorbox}
\caption{Few-shot prompt used for GPT-4o-mini rule categorization. The top section lists the 20 possible categories; the middle section shows the 12 labeled examples; the bottom section shows the final task query with the target rule substituted in.}
\label{fig:few-shot-prompt}
\end{figure}

In~\ref{fig:few-shot-prompt} we show the prompt that was used for categorization. In our final prompt, we use twelve examples randomly sampled across instances of varying sizes, rule types, and combinations of rule types. 
We incorporate a strict Q-A style of prompt to ensure that our final output in consistent and thereby easy to parse and aggregate.

Prior to using this prompt for categorizing our entire rule dataset, we experimented with varying $n$ for few-shot prompting, addition of description of rule types, and model types. 
We experimented with different number of prompts: eight, twelve and sixteen. We found that eight or sixteen did not give any substantial gains.  
Prior to experimenting with few-shot prompting, we only provided rule types and definitions. However, we found that in this setup, LLM starts creating it's own categories and using varied naming conventions that made it hard to assimilate results.
The addition of rule description to the few-shot prompting did not improve its performance. 
Lastly, we tested across two models: GPT-3.5-turbo and GPT-4o-mini, with temperature setting at the default 1.0. 
To ensure consistency, we ran the few shot prompt~\ref{fig:few-shot-prompt} three times with GPT-3.5-turbo and GPT-4o-mini. GPT-4o-mini consistently outperformed GPT-3.5-turbo, though by just 1-2\% points. Each run with GPT-4o-mini yielded an accuracy between 75-76\%. Ultimately, we used the best performing run and report that in the main paper~\ref{Rule Categorization}.

\section{Robustness of Rule Topics as Community Topical Focus Measures}
\label{appx:topical_focus}

We investigate the focus of communities by using the topics inferred from rules. In this section, we examine an alternative operationalization that uses community description. 
We used the `Short Description' endpoint from the Mastodon API to extract the topic of an instance. To categorize these topics, we adapted a set of topics derived from an analysis of 5.1 billion Reddit comments between 2005 and 2018 \cite{waller_quantifying_2021}. The first author conducted inductive coding to assess its suitability for our study, finding that the Reddit topic set provides good coverage for topics encountered on Mastodon.
Two authors independently annotated 100 randomly sampled instance descriptions in the first round using this topic set. They then refined the taxonomy through discussion, producing a comprehensive list of topics and definitions (see \appendixautorefname{}, Table \ref{instance_topic_taxonomy}). A second round of annotation was conducted on another 100 instance descriptions to calculate an inter-rater reliability score. Using Krippendorff's alpha, which accounts for multiple labels \cite{krippendorff2004}, we obtained a score of 0.79, indicating high agreement between annotators.

We then use GPT-4o to automatically annotate the full dataset of 6,660 instance short descriptions. The prompt consisted of the taxonomy definitions (as described in \appendixautorefname{}, Table \ref{instance_topic_taxonomy}), and a few randomly sampled examples. 
To evaluate performance, we tested GPT-4o on the same 100-sample set coded by the authors, which yielded high agreement with human annotations ($\alpha=0.71$ and 0.69). This result indicates that the prompt enabled GPT-4o to generate high-quality annotations.


\begin{longtable}{m{1cm} m{3cm} m{9cm}}
\caption{Taxonomy for the topics of Mastodon instances} \\
\hline
\textbf{ID} & \textbf{Topic} & \textbf{Description} \\
\hline
\endfirsthead

\hline
\textbf{ID} & \textbf{Topic} & \textbf{Description} \\
\hline
\endhead

1  & General          & Instances that range in a wide variety of topics \\
\hline
2  & Family \& Friends & Instances that explicitly mention that it is only for friends and family; or invite-only for people in the know \\
\hline
3  & Personal         & Instances that explicitly mention that it is a personal instance \\
\hline
4  & Gaming           & Instances that are about video games \\
\hline
5  & Tech             & Instances about various tech topics such as coding, programming, open source etc. \\
\hline
6  & Fediverse        & Instances are coming from other Fediverse platforms such as Pleroma or Pixelfed \\
\hline
7  & Geographic       & Instances that are specifically for people of some countries or specific cities \\
\hline
9  & Interests        & Instances that are for niche interests that may not fit into other categories \\
\hline
8  & Ideology         & Instances that specify focus on ideologies (such as equality, freedom, etc.) \\
\hline
10 & Entertainment    & Instances that are about books, TV, shows, fandoms, music etc. \\
\hline
11 & Sexual content   & Instances that are about sexual material including but not limited to photo-sharing and sexual role-play \\
\hline
13 & LGBTQ            & Instances that explicitly state they are for people of the LGBTQ community \\
\hline
14 & News             & Instances that focus on conversations about news \\
\hline
16 & Art \& Culture   & Instances about the creation and sharing of art \& culture \\
\hline
17 & Self-improvement & Instances for people to talk about ways to improve themselves and the world around them \\
\hline
18 & Activism         & Instances for promoting activism related to more politically charged topics \\
\hline
19 & Organization     & Instances for organizations such as companies, schools/universities, club chapter \\
\hline
20 & Support          & Instances that focus on safety and inclusivity \\
\hline
22 & Politics         & Instances to talk about politics and current affairs \\
\hline
23 & Bots             & Instances specifically for bots \\
\hline
24 & Religion         & Instance focused on conversation about faith \\
\hline
26 & Pets             & Instances to talk about all sorts of pets such as dogs, cats etc. \\
\hline
27 & N/A              & Instances that don’t specify purpose or target audience \\
\hline
28 & Memes            & Instances about the sharing of memes and deep discussion about its cultural relevance \\
\hline
\end{longtable}
\label{instance_topic_taxonomy}

\begin{figure}
\centering
\includegraphics[width=\textwidth]{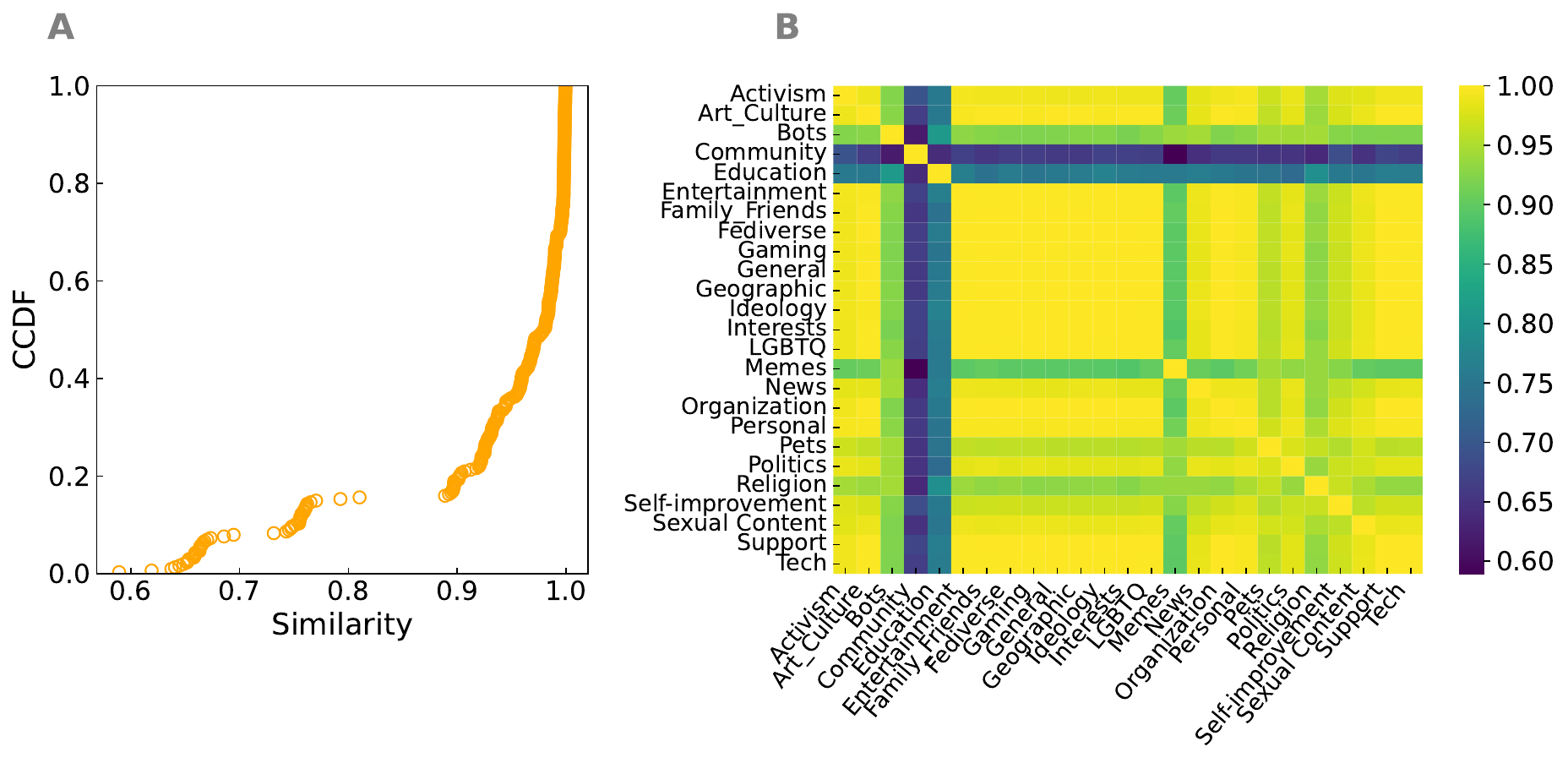}
\caption{(a) CCDF and (b) heatmap of similarities between topics based on their rule categories}
    \label{fig:topic-category-ccdf}
\end{figure}

We calculate the similarity between instance description topics based on their distributions of rule categories. Each topic is represented as a 19-dimensional vector, where each dimension corresponds to the frequency of a rule category within that topic. We normalize these vectors using TF-IDF to reduce the influence of commonly occurring categories. This ensures that topics sharing widely prevalent categories are not falsely considered to be similar. Finally, we compute the cosine similarity between the normalized vectors. A cosine similarity of 1 indicates that two topics have identical distributions of rule categories, while 0 means they share no categories in common.

\section{Effects of Twitter/X Migration on Instance Rule Sets}
\label{appx:wayback}

Wayback Machine is an initiative to create an archive of the Internet. Previous research has used this archive to capture historical data \cite{reddy_evolution_2023, Leibmann_Weld_Zhang_Althoff_2025}. 
We queried the CDX API\footnote{\url{github.com/internetarchive/wayback/tree/master/wayback-cdx-server}} to collect instance rule sets from October 2022, January 2023, and January 2024. These timepoints correspond to the period before Twitter/X's acquisition on October 27, 2022, and approximately one month and one year after the acquisition, respectively.

We began by collecting data from a list of popular instances compiled by Nicholson et al. (\citeyear{nicholson_mastodon_2023}). Only 10\% of these 100 instances had snapshots available in the Wayback Machine prior to October 2022. Because most Personal instances are single-use, we did not expect substantial changes or consistent archival by the Wayback Machine. Thus, to expand coverage while maintaining relevance, we collected snapshots across multiple timestamps for all instances not tagged as Personal (N=4,809, around 13\%). Of these, only 130 instances had rule sets prior to Oct 2022 that could be recovered from Wayback Machine, and only 60 instances contained rule sets across all three timestamps of interest---enabling longitudinal comparison. 

To analyze changes across periods 2022--2023 and 2023--2024, we measure differences in user count, rule count, word count and rule topic categories. For user count, we record the number of users per instance at each timestamp and calculated differences between January 2023 and October 2022, and between January 2024 and January 2023. We average these differences across instances and report the mean and standard deviation (Table~\ref{table:rule-change-migration}). Rule count is measured similarly, based on the number of rules per instance. For word count, we calculate the total number of words in each rule set, again reporting average differences and standard deviations across instances.

For changes in rule topic categories, we apply the prompt described in Section~\ref{Rule Categorization} to classify rules at each timestamp. We then compare rule sets using Jaccard similarity, where 0 indicates no commonality, and 1 indicates exact match. We use a lax threshold of 0.9 to distinguish stability from change: rule sets with similarity above 0.9 were considered unchanged, while those below the threshold were treated as different. Yet, only 11 of 60 instances (19.05\%) changed between 2022–2023, and 5 of 60 (9.52\%) changed between 2023--2024 under this criterion.


\begin{table}[ht]
\centering
\begin{tabular}{lrrrrrrc}
\hline
\textbf{Transition} &
\multicolumn{2}{c}{\textbf{User count change}} &
\multicolumn{2}{c}{\textbf{Rule count change}} &
\multicolumn{2}{c}{\textbf{Word count change}} &
\textbf{Number of instances} \\
\cline{2-8}
& \textbf{Mean avg.} & \textbf{Std} & \textbf{Mean avg.} & \textbf{Std} & \textbf{Mean avg.} & \textbf{Std} &  \\
\hline
2022 $\to$ 2023 & 118.77  & 787.08  & 0.07   & 0.52 & 0.54  & 4.17 & 11\\
2023 $\to$ 2024 & 1843.09 & 5738.96 & -0.03  & 0.26 & -0.34 & 5.84 & 5\\
\hline
\end{tabular}
\caption{Statistics of user count, rule count, word count, and rule type corresponding to different transition time points.}
\label{table:rule-change-migration}
\end{table}

\end{document}

%% file: CSCW/table_primary.tex
\begin{table}[ht]
\centering
\caption{Size and Federation Effects on Rule Formalization}
\label{table:nb_regression}

\begin{tabular}{lccccc}

\cline{1-4} \cline{5-5} \cline{6-6}
\vspace{-0.5em} \\
\textbf{Predictors} & \textbf{Word Count$^{\dagger}$} & \textbf{Rule Count$^{\dagger}$} & \textbf{Topic Count$^{\dagger}$} & \textbf{TTR$^{\ddagger}$} & \textbf{Readability$^{\S}$} \\
\hline
\vspace{-0.5em} \\
Log User Count  & 0.42*** & 0.29*** & 0.20*** & $-$0.47*** & $-$8.14*** \\
                & (0.02)  & (0.01)  & (0.01)  & (0.02)     & (0.56)     \\
\vspace{-0.5em} \\
Log Fed Num     & 0.05    & 0.08*** & 0.06*   & $-$0.11*** & 2.35*      \\
                & (0.03)  & (0.02)  & (0.03)  & (0.03)     & (1.07)     \\
\vspace{-0.5em} \\
Intercept       & 3.41*** & 0.82*** & 1.04*** & 2.74*** & 65.10*** \\
                & (0.11)  & (0.07)  & (0.10)  & (0.11)  & (4.07)   \\
\hline
\vspace{-0.5em} \\
Observations        & 4,801  & 4,801  & 3,621 & 4,363  & 4,363  \\
Pseudo-$R^2$        & 0.015  & 0.036  & 0.018        & ---    & 0.047  \\
AIC                 & 49,128 & 23,533 & 18,504       & $-$12,181 & 43,486 \\

\hline
\end{tabular}%

\vspace{0.5em}
\begin{flushleft}
\footnotesize
\textit{Note:} Unstandardized coefficients shown. Standard errors in parentheses (HC3 robust for OLS). \\
*** $p < 0.001$, ** $p < 0.01$, * $p < 0.05$. \\
$^{\dagger}$ Negative Binomial regression. Coefficients represent the log of expected counts. \\ 
$^{\ddagger}$ Beta regression. \\
$^{\S}$ OLS 
\end{flushleft}
\end{table}

%% file: CSCW/table.tex

\begin{table}[ht]
\centering
\caption{Full Models with Age Controls}
\label{table:robust_regression}

\begin{tabular}{lccccc}
\hline
\vspace{-0.5em} \\
\textbf{Predictors} & \textbf{Word Count$^{\dagger}$} & \textbf{Rule Count$^{\dagger}$} & \textbf{Topic Count$^{\dagger}$} & \textbf{TTR}$^{\ddagger}$ & \textbf{Readability$^{\S}$}\\
\hline
\vspace{-0.5em} \\
Log User Count  & 0.29*** & 0.20*** & 0.20*** & $-$0.36*** & $-$7.83*** \\
                & (0.02)  & (0.01)  & (0.01)  & (0.02)     & (0.56)     \\
\vspace{-0.5em} \\
Log Fed Num     & 0.02    & 0.05*   & 0.08**  & $-$0.14*** & 1.97       \\
                & (0.03)  & (0.02)  & (0.02)  & (0.04)     & (1.12)     \\
\vspace{-0.5em} \\
Age (years)     & $-$0.03*** & $-$0.02*** & $-$0.04*** & 0.05*** & 0.87*** \\
                & (0.01)     & (0.01)     & (0.01)     & (0.01)  & (0.22)  \\
\vspace{-0.5em} \\
Intercept       & 3.99*** & 1.31*** & 1.18*** & 2.34*** & 62.45*** \\
                & (0.13)  & (0.08)  & (0.09)  & (0.15)  & (4.24)   \\
\hline
\vspace{-0.5em} \\
Observations        & 3,356  & 3,356  & 3,356  & 3,059 & 4,357  \\
$R^2$ / Pseudo-$R^2$ & 0.009 & 0.025  & 0.023  & ---          & 0.045  \\
AIC                 & 36,056 & 16,965 & 16,919 & $-$4,712     & 43,390 \\
\hline

\end{tabular}%

\vspace{0.5em}
\begin{flushleft}
\footnotesize
\textit{Note:} Unstandardized coefficients shown. Standard errors in parentheses (HC3 robust for OLS). \\
*** $p < 0.001$, ** $p < 0.01$, * $p < 0.05$. \\
$^{\dagger}$ Negative Binomial coefficients represent the log of expected counts. \\
$^{\ddagger}$ Beta regression model\\
$^{\S}$ OLS 
\end{flushleft}
\end{table}
